\documentclass[12pt,preprint]{aastex} 
\shorttitle{Components of $\delta$ Ori A} 
\shortauthors{Richardson et al.} 
 
\begin{document} 
 
\received{} 
\accepted{} 
 
\title{HST/STIS Ultraviolet Spectroscopy of the Components 
of the Massive Triple Star $\delta$~Ori~A\footnote{Based on 
observations made with the NASA/ESA Hubble Space Telescope, 
obtained at the Space Telescope Science Institute, which is operated by the 
Association of Universities for Research in Astronomy, Inc., under NASA contract 
NAS 5-26555. These observations are associated with program \#13450.}}  
 
\author{Noel D. Richardson, Anthony F. J. Moffat}
\affil{D\'epartement de physique and Centre de Recherche en Astrophysique du Qu\'ebec (CRAQ), 
 Universit\'e de Montr\'eal, C.P. 6128, Succ.~Centre-Ville, Montr\'eal, Qu\'ebec, H3C 3J7, Canada; \\
 richardson@astro.umontreal.ca, moffat@astro.umontreal.ca}

\author{Theodore R. Gull, Don J. Lindler}
\affil{Astrophysics Science Division, NASA Goddard Space Flight Center, 
 Greenbelt, MD 20771, USA; \\
 theodore.r.gull@nasa.gov, don.j.lindler@nasa.gov}

\author{Douglas R. Gies} 
\affil{Center for High Angular Resolution Astronomy and  
 Department of Physics and Astronomy,\\ 
 Georgia State University, P. O. Box 5060, Atlanta, GA 30302-5060, USA; \\ 
 gies@chara.gsu.edu} 
 
\author{Michael F. Corcoran}
\affil{CRESST and X-ray Astrophysics Laboratory NASA/GSFC, Greenbelt, MD 20771, USA; 
 Universities Space Research Association, 7187 Columbia Gateway Drive, Columbia, MD 21046, USA; \\
 michael.f.corcoran@nasa.gov}

\author{Andr\'{e}-Nicolas Chen\'{e}}
\affil{Gemini Observatory, Northern Operations Center, 670 North A'ohoku Place, Hilo, HI 96720, USA; \\
 achene@gemini.edu}

\slugcomment{Accepted to ApJ} 
\paperid{99022}

 
\begin{abstract} 
The multiple star system of $\delta$~Orionis is one of the closest
examples of a system containing a luminous O-type, bright giant star
(component Aa1).  It is often used as a spectral-type standard and 
has the highest observed X-ray flux of any hot-star binary.
The main component Aa1 is orbited by two lower mass stars, faint Aa2 
in a 5.7 day eclipsing binary, and Ab, an astrometric companion with 
an estimated period of 346 years.  Generally the flux from all three 
stars is recorded in ground-based spectroscopy, and the spectral 
decomposition of the components has proved difficult.  Here we present 
HST/STIS ultraviolet spectroscopy of $\delta$~Ori~A that provides 
us with spatially separated spectra of Aa and Ab for the first time. 
We measured radial velocities for Aa1 and Ab in two observations 
made near the velocity extrema of Aa1.  We show tentative evidence
for the detection of the Aa2 component in cross-correlation functions 
of the observed and model spectra.  We discuss the appearance of the 
UV spectra of Aa1 and Ab with reference to model spectra.  
Both stars have similar effective temperatures, but Ab is fainter 
and is a rapid rotator.  The results will help in the interpretation 
of ground-based spectroscopy and in understanding the physical and 
evolutionary parameters of these massive stars. 
\end{abstract} 
 
 
 
\setcounter{footnote}{0} 
 
\section{Introduction}                              

The western-most of the three stars in the ``Belt of Orion'' is the 
multiple star system $\delta$~Orionis (Mintaka; HD 36486), and the 
brightest component of the system represents one of the nearest
of the luminous O-type stars (O9.5~II~Nwk; Sota et al.\ 2011). 
The system consists of a close eclipsing pair Aa1,Aa2 that is 
orbited by Ab with a current angular separation of $0\farcs3$
(Tokovinin et al.\ 2014), and there are two other more distant 
components B and C (Harvin et al.\ 2002).  The system was the
target of a recent multiwavelength observing campaign to explore 
its X-ray properties (Corcoran et al.\ 2015; Nichols et al.\ 2015),
photometric and spectroscopic variability (Pablo et al.\ 2015), and 
spectral properties (Shenar et al.\ 2015).  These studies depend
on stellar parameters of the O-star and its close companions, 
and there remains some confusion about the nature of the companions
because the flux of all three stars is generally recorded in 
observations with limited angular resolution.  The single-lined 
spectroscopic orbit of the O-star (component Aa1) is well established 
(Harvey et al.\ 1987), but attempts to find and measure the Doppler 
shifts of its companion Aa2 have led to different conclusions. 
Harvin et al.\ (2002) analyzed the available collection of 
ultraviolet spectra from the {\it International Ultraviolet Explorer}
satellite, and they fit cross-correlation functions of the 
observed spectra with that of a sharp-lined template star 
to determine a double-lined solution that led to relatively 
small mass estimates.  Mayer et al.\ (2010) presented a study of
the light curve and high S/N optical spectra, and they argued 
that the spectral signature of the Ab component was significant 
while that from Aa2 was too faint to detect.  They suggested 
that the light and radial velocity curves of the Aa1,Aa2 binary 
were consistent with the masses expected for their spectral 
classifications.  Harmanec et al.\ (2013) extended this work 
to search for evidence of the Aa2 spectral component in a 
large set of spectra recording the \ion{He}{1} $\lambda 6678$ 
absorption line, and their spectral reconstruction methods
detected the faint companion signal with an orbital semiamplitude
of $K_2 = 273$ km~s$^{-1}$ (leading to a mass ratio $q=0.40$). 
The conclusions of Mayer et al.\ and Harmanec et al.\ are generally 
verified in the recent orbital analysis by Pablo et al.\ (2015) and 
the spectral investigation by Shenar et al.\ (2015).  
The stellar properties of $\delta$~Ori are consistent with its membership 
in the Ori~OB1b association (at a distance of 380~pc for the nearby 
$\sigma$~Ori cluster; Caballero \& Solano 2008), which is about 
twice as far as estimated by the {\it Hipparcos} parallax 
(Shenar et al.\ 2015). 

The most direct solution to the problem of the spectral identities
of the stars is to obtain angularly resolved spectroscopy of the 
Aa and Ab components separately.  Here we present the first direct
resolution of the spectral components through ultraviolet (UV) spectroscopy 
with the Space Telescope Imaging Spectrograph (STIS) aboard the 
{\it Hubble Space Telescope}.  We describe in \S2 the STIS 
observations made during two visits that occurred near the 
times of the velocity extrema of the Aa1,Aa2 binary system. 
We used cross-correlation methods to measure the radial velocities
of Aa1 and Ab that we present in \S3.  Average versions of the 
UV spectra of Aa1 and Ab are compared with model spectra 
in \S4 in order to estimate their effective temperatures 
and projected rotational velocities.  We summarize our results 
in \S5 and discuss how they may be used to analyze spectra 
that record the flux of all three stars. 

 
\section{Spectroscopic Observations}                

The STIS UV spectra of $\delta$~Ori were obtained during visits on 
2014 Dec 30 and 2015 Feb 10, which correspond to times near Aa1 
radial velocity maximum and minimum, respectively, based upon 
the ephemeris of Mayer et al.\ (2010).  We used the 
E140M echelle grating that records the UV spectrum over the 
range 1144 to 1730 \AA\  with a spectral resolving power of 
$R = \lambda/\triangle\lambda \sim 46000$ 
(Kimble et al.\ 1998; Biretta et al.\ 2015). 
The 0.3X0.05ND slit was oriented with the long axis approximately 
perpendicular to the position angle between Aa and Ab in order
to minimize any flux contamination from the other star at $0\farcs3$ 
separation. 
The alternative of placing both stars on the slit is impractical 
because of the close spacing of the echelle orders on the MAMA detector.
We expect that at this separation the intensity from the respective 
companion will be reduced by about $400\times$ its peak value 
according to the spatial point spread function (PSF) for the FUV-MAMA
detector (Biretta et al.\ 2015).  A first exposure of Aa of 300~s 
duration was made before moving to Ab for a longer exposure of 
1360~s.  The spatial profile of the Aa spectrum appeared 
double-peaked in data from the first visit, while the 
spatial distribution of the Ab spectrum appeared double-peaked 
in the data from the second visit.  We suspect that this 
anomaly resulted from tracking errors related to the guide stars
used in these observations.  This has no impact on the 
wavelength calibrations and relative spectral fluxes of the 
individual observations.  Note that the narrow width of the aperture
subsamples the PSF of {\it HST}, and small variations due to 
thermal shifts of focus contribute to changes in the flux throughput. 
Thus, our observations are not well suited to estimate the 
absolute flux calibrations of the spectra.  For example, 
we find that the mean flux ratio in the region near 1319 \AA\ is 
$F_\lambda ({\rm Ab}) / F_\lambda ({\rm Aa}) = 0. 38$ and 0.26 
for the first and second visits, respectively, which is a much
larger variation than expected for the orbital light curve 
(Pablo et al.\ 2015). 

The STIS spectra were extracted and calibrated using the 
standard STScI pipeline (Bostroem \& Proffitt 2011).
The final product is a {\tt x1d.fits} file of absolute flux 
as a function of heliocentric wavelength for each echelle order. 
We transformed these onto a standard wavelength grid in 
increments of $\log\lambda$ equivalent to a pixel step of 
10 km~s$^{-1}$.  This was accomplished by forming the error weighted 
mean for each wavelength bin from all the available input data 
(including points from adjacent orders).  This resulted in a 
slight degradation of the resolving power but an increase in 
the net signal-to-noise ratio, reaching a value of $S/N \approx 30$
per pixel in the best exposed parts of the spectrum.  There were some small 
gaps in wavelength coverage between a few orders long-ward of 1680 \AA , 
and the spectrum was interpolated in wavelength for those missing wavelength bins. 
Each spectrum was normalized by a second order fit of the fluxes. 
The UV spectrum of $\delta$~Ori is inscribed with a rich collection 
of interstellar lines, and most of these were removed 
by direct interpolation across the features so that their 
impact is minimal in the following treatment of the stellar spectra. 

 
\setcounter{footnote}{1} 
 
\section{Radial Velocities and Orbital Elements}    

We measured radial velocities for each of the Aa1 and Ab spectral 
components by calculating the cross-correlation function (CCF) 
of the observed spectrum with a model template.   The models
were formed by interpolation in the OSTAR2002 grid of synthetic 
spectra\footnote{http://nova.astro.umd.edu/}  
from Lanz \& Hubeny (2003).  These models 
are based upon line-blanketed, non-LTE stellar atmospheres for 
static photospheres, so they do not include the wind features 
found in luminous star spectra.  In fact, the wind lines 
display minimal orbital motion because much of the line formation
occurs far from the star where the motion is primarily radial, 
so the wind line regions were excluded in the CCF calculation
(omitting the entire low wavelength range 1150 -- 1249 \AA\ and the 
regions surrounding \ion{Si}{4} $\lambda 1400$ and \ion{C}{4} $\lambda 1550$).
The rotationally broadened model spectra were selected using the stellar 
parameters for Aa1 and Ab derived by Shenar et al.\ (2015).  The resulting 
CCFs for Aa1 and Ab are illustrated in the left and right panels, 
respectively, of Figure~1.  Component Ab appears to be radial velocity 
constant between the observations, as expected for its long orbital 
period ($P = 346$~y; Tokovinin et al.\ 2014).  However, component 
Aa1 shows the Doppler shifts corresponding to its orbital motion in the close binary.  
Noteworthy by its absence is any indication of a signal in the 
CCF from the Aa2 companion, which is probably due to its relative 
faintness.  We derived radial velocities by fitting a parabola 
to the peak of the CCF, and these velocities are listed in Table~1.  
Table~1 gives the component identification, 
the heliocentric Julian data at the time of mid-exposure, the 
orbital phase relative to phase zero at eclipse minimum 
(Pablo et al.\ 2015), the radial velocity, and its uncertainty 
(from analytic estimates using the method of Zucker 2003).

\placetable{tab1}      
  
\begin{figure}
\begin{center} 
\plottwo{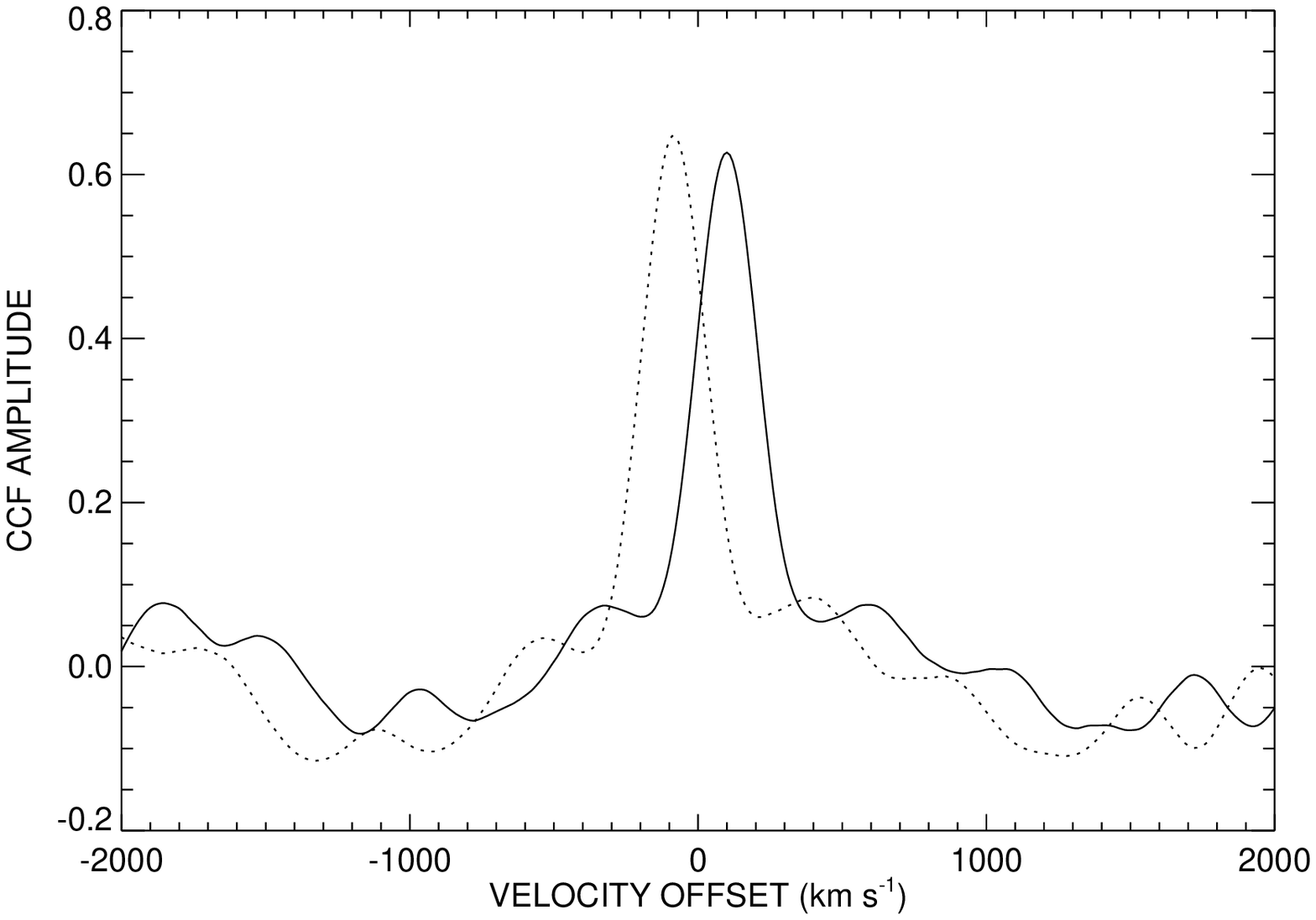}{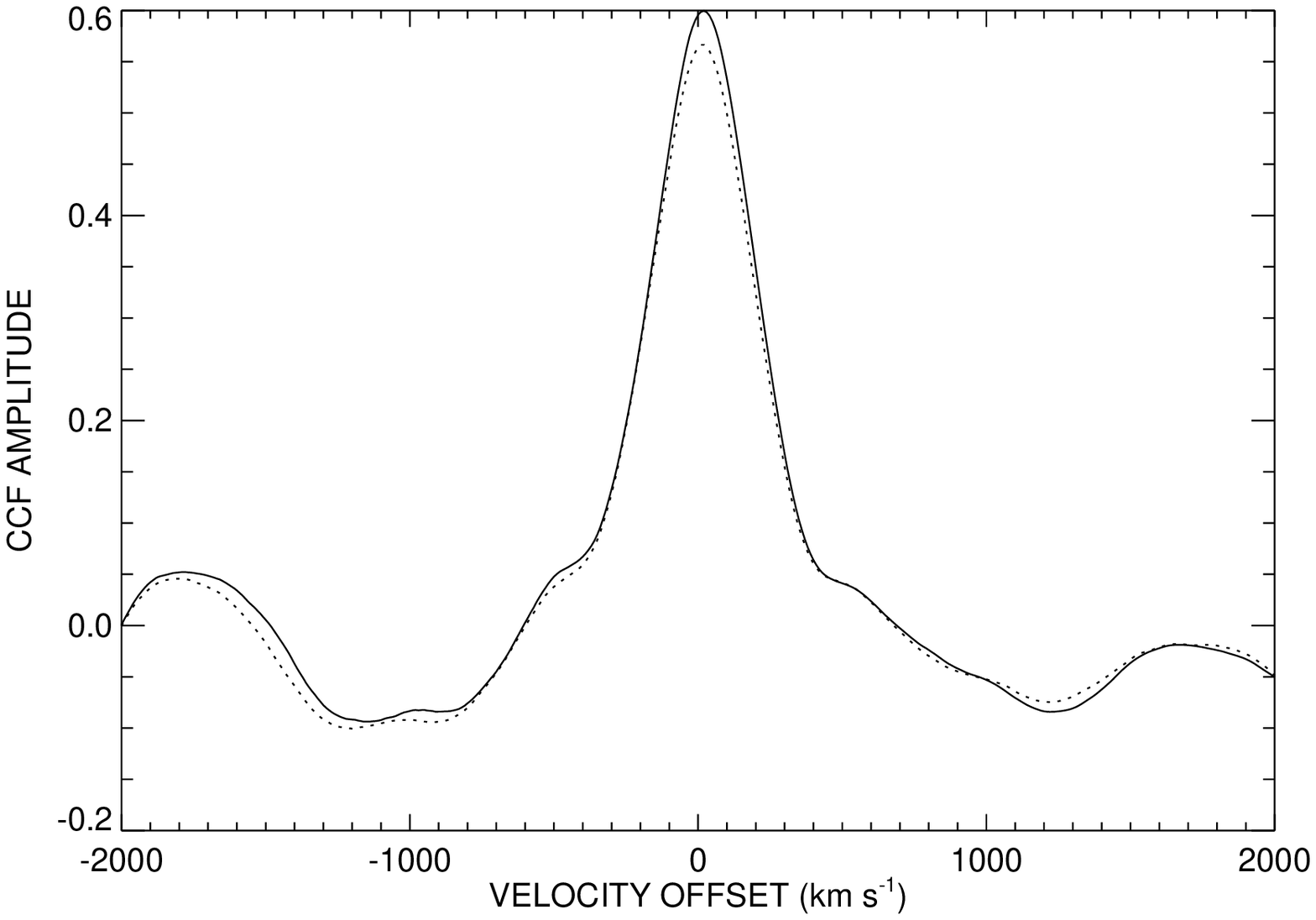}
\end{center} 
\caption{The cross-correlation functions for $\delta$~Ori~Aa1
(left) and $\delta$~Ori~Ab (right).  The solid and dotted lines
show the results for the first and second STIS observation, 
respectively.   
\label{fig1}} 
\end{figure}

\placefigure{fig1}     

The lack of evidence of Aa2 in the CCFs presented in Figure~1 
clearly indicates that it contributes only a small fraction of 
the UV flux.  We attempted a search for its spectral contribution 
by considering the difference spectrum of our two quadrature 
observations of Aa.  Both flux normalized spectra of Aa were shifted 
to the rest frame of Aa1 using the measured velocity shifts in Table~1, 
and then the second spectrum was subtracted from the first to 
remove most of the contribution of Aa1.  The resulting difference
spectrum will also be equal to the residual spectrum of Aa2 at its 
blueshifted position in the first spectrum minus that for its
redshifted position in the second spectrum.  Thus, we might expect 
that the CCF of this difference spectrum would display a blueshifted 
peak (correlation with the absorption lines in the first spectrum) 
and a redshifted trough (anticorrelation with the reversed absorption
lines in the second spectrum).  
We calculated several model template spectra for Aa2 using 
stellar parameters close to those found by Shenar et al.\ (2015) 
with models from the BSTAR2006 grid of Lanz \& Hubeny (2007), 
which cover the lower temperature range expected for Aa2. 
The best correlation was obtained with 
selected values of $T_{\rm eff} = 23$~kK, $\log g = 4.0$, and 
$V\sin i = 40$ km~s$^{-1}$.  The latter was chosen from the 
widths of the peaks in the CCF when a $V\sin i = 0$ model was 
used as a template.   
The projected rotational velocity of Aa2 is smaller than that found 
by Shenar et al.\, $V\sin i = 150 \pm 50$ km~s$^{-1}$,
but the disagreement is not surprising given that Shenar et al.\
obtained their measurement from a difficult blend analysis of the weak 
line wings of \ion{He}{1} $\lambda\lambda 4026,4144$ (see their Fig.~4).
The CCF of the difference spectrum and the 
model is shown in Figure~2.   This CCF is clearly dominated by 
noise in the difference spectrum, but there does appear to be 
a peak and a trough at the velocities relative to the Aa1 frame
that should correspond to the Doppler shifts of Aa2.  There is 
also a weaker peak at zero velocity that is associated with the
lingering presence of the features of Aa1 in the difference spectrum.
We made parabolic fits of the positions of the peak and trough 
extremities (indicated by the dotted lines in Fig.~2), and these
velocities were added to the Aa1 velocities to yield the heliocentric
system velocities reported in Table~1.  Because this is a tentative 
detection, we simply estimated the uncertainties as $0.5\times$ HWHM 
of the peak and trough distributions in the CCF. 

\begin{figure}
\begin{center} 
{\includegraphics[angle=90,height=12cm]{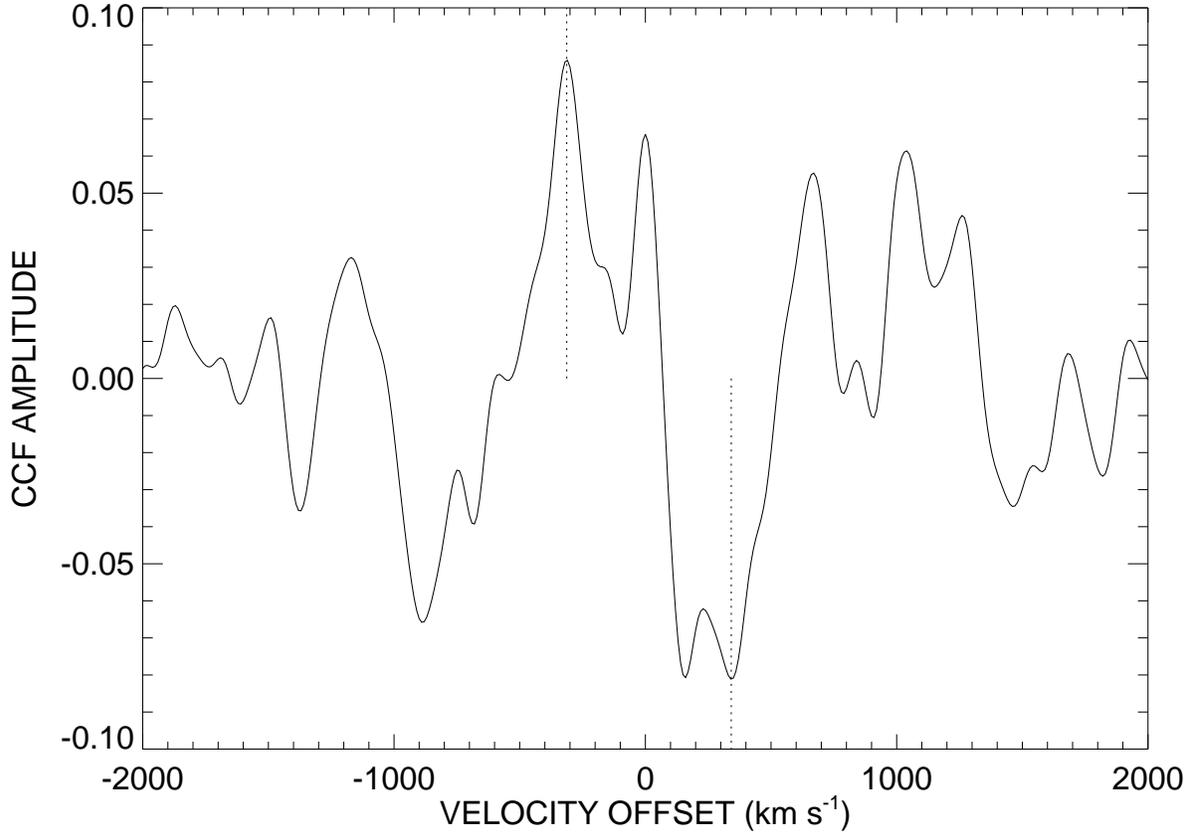}}
\end{center} 
\caption{The cross-correlation function of the difference
of the two $\delta$~Ori~Aa spectra in the reference frame of
component Aa1. A model spectrum for Aa2 was used as the template. 
The vertical dotted lines indicate the possible detection of 
Aa2 in the first (left) and second (right) observation.
These velocity offsets correspond to the predicted shifts of Aa2
at the times of the observations.
\label{fig2}} 
\end{figure}

\placefigure{fig2}     

We can compare these velocity measurements for Aa1 with the recent 
orbital solution from a large set of contemporaneous, ground-based 
spectroscopy presented by Pablo et al.\ (2015).  We began by assuming 
the period $P$ from the work of Mayer et al.\ (2010) and the 
epoch of periastron $T$, eccentricity $e$, and longitude of 
periastron $\omega$ from the ``LM'' (low mass) solution of 
Pablo et al.\ (2015).   We then adopted the rate of periastron 
advance derived by Pablo et al.\ to prorate $T$ and $\omega$ to 
the periastron that occurred at the mean time of our observations
(130 orbital cycles after the periastron epoch given by Pablo et al.). 
We are then left with two radial velocities to find the remaining 
two orbital elements, the systemic velocity $\gamma$
and the semiamplitude $K$.  In such a situation, we cannot estimate
parameter uncertainties from the residuals to the fit, because the 
residuals are zero, so we simply set the errors in these two 
parameters equal to the mean error of the velocities. 
We used the program described by Morbey \& Brosterhus (1974) 
to find the values of $\gamma_1$, $\gamma_2$, $K_1$, and $K_2$ that
are listed in Table~2.  Our estimate of $\gamma_1$ is intermediate 
between those determined by Mayer et al.\ (2010) and Pablo et al.\ (2015), 
and small differences in systemic velocity are not unexpected 
given the different methods of radial velocity measurement. 
Our derived value of $K_1 = 104.6 \pm 1.6$ km~s$^{-1}$ falls between 
those found by Pablo et al.\ (2015) and Harmanec et al.\ (2013), 
$K_1 = 96.0 \pm 0.6$ and 109.0 km~s$^{-1}$, respectively. 
Our estimate of $K_2 = 266 \pm 20$ km~s$^{-1}$ agrees within uncertainties
with that found by Harmanec et al.\ (2013), $K_2 = 273$ km~s$^{-1}$.

\placetable{tab2}      
  
The mass ratio we derive from our tentative detection of the CCF
signal of Aa2, $q = M_2/M_1 = 0.39 \pm 0.03$, is consistent with 
estimates from the eclipsing light curve (Luyton et al.\ 1939; 
Koch \& Hrivnak 1982; Mayer et al.\ 2010; Pablo et al.\ 2015).  
Table~2 lists the resulting masses for Aa1 and Aa2 that we 
derive using an orbital inclination $i=76\fdg3$ from Pablo et al.\ (2015).
Both stars have masses in line with expectations from their 
temperatures and luminosities 
(Martins et al.\ 2005; Torres et al.\ 2010), 
and our work confirms that 
the lower masses found by Harvin et al.\ (2002) resulted from an
incomplete treatment of the spectral signature of component Ab.
The mass of the entire triple system derived from the preliminary astrometric 
orbit (Tokovinin et al.\ 2014) is $40 M_\odot (d/400 ~{\rm pc})^3$
where $d$ is the distance to the Ori~OB1b association in which 
$\delta$~Ori resides.  
The difference between the total mass and 
the Aa masses in Table~2 would imply that the mass of Ab is only 
$8 M_\odot$, much lower than the $22 M_\odot$ mass estimated by 
Shenar et al.\ (2015) for a star of its effective temperature 
and luminosity.   We suspect that future revisions to 
the astrometric orbit of Aa,Ab may well change this low mass estimate. 

 
\section{Spectral Properties}                       
 
The excellent quality of the STIS spectra of components Aa1 and Ab 
encouraged us to compare their spectral morphology to model spectra 
in order to make independent estimates of the stellar parameters. 
We formed a simple shift-and-add average of the Ab spectra 
to move them to the rest frame.  We did the same for the Aa1 
spectra (using the velocities in Table~1), but the results 
are marred in the vicinity of the wind lines because they 
do not exhibit orbital motion.  Instead, we made a simple mean
spectrum (shifted to the rest frame using $\gamma_1$) to 
represent the spectrum of Aa1 in spectral regions near the wind 
features of \ion{C}{3} $\lambda 1175$, Ly$\alpha$ $\lambda 1216$,
\ion{N}{5} $\lambda 1240$, \ion{Si}{4} $\lambda 1400$, and 
\ion{C}{4} $\lambda 1550$.  Figure~3 shows the two versions 
of the Aa1 spectrum using black plus signs for the shift-and-add 
average of the photospheric parts and using gray crosses for the 
simple average applied to the wind line regions. 
Figure~4 likewise illustrates the average spectrum of component Ab.
Both figures are marked by small vertical ticks near the 
bottom that indicate the positions of the interstellar lines 
that were removed.  
The spectra of Aa1 and Ab shown in Figures 3 
and 4 are available as FITS files attached to the electronic 
version of this paper.

\begin{figure} 
\begin{center} 
{\includegraphics[angle=90,height=12cm]{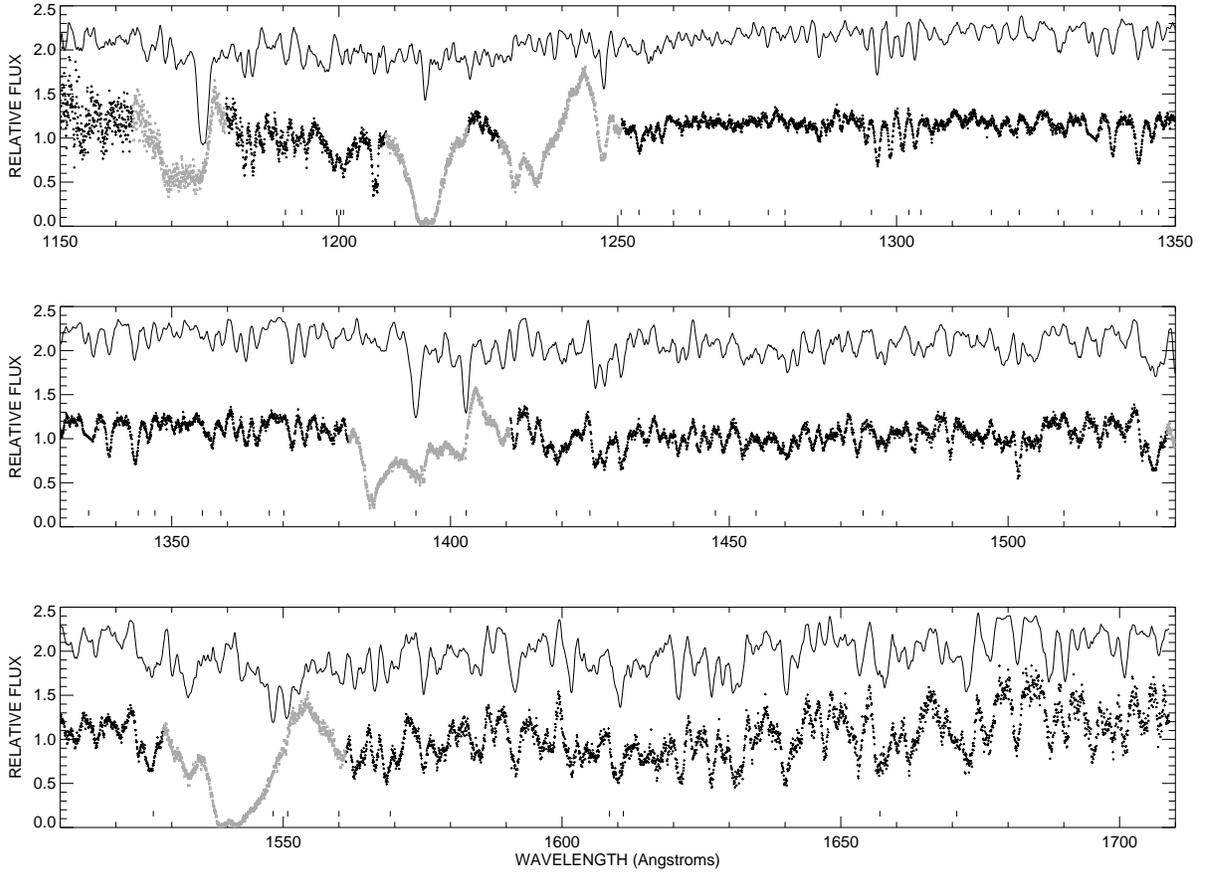}} 
\end{center} 
\caption{The normalized ultraviolet spectrum of $\delta$~Ori~Aa1 (below) 
compared with a model from TLUSTY (solid line above, offset by 0.8 for clarity). 
The black plus signs indicate the result of shifting and adding the spectrum
to the reference frame of Aa1, while the gray crosses show a simple average
spectrum in the rest frame in the vicinity of the stellar wind lines.
The lower line segments indicate the locations 
where interstellar lines were removed from the spectrum. 
\label{fig3}} 
\end{figure}

\placefigure{fig3}     

\begin{figure} 
\begin{center} 
{\includegraphics[angle=90,height=12cm]{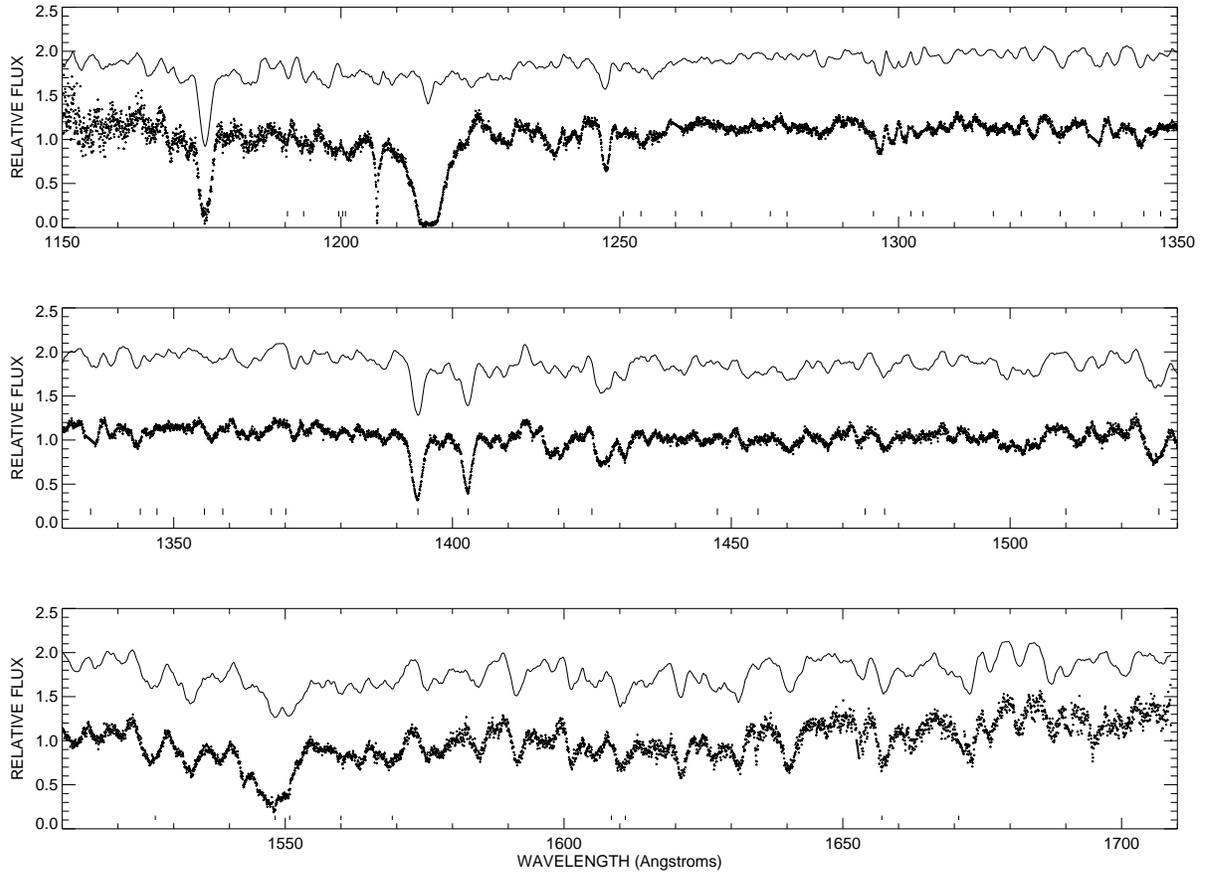}} 
\end{center} 
\caption{The normalized ultraviolet spectrum of $\delta$~Ori~Ab (plus signs) 
compared with a model from TLUSTY (solid line, offset by 0.8 for clarity)
in the same format as Figure~3.
\label{fig4}} 
\end{figure}

\placefigure{fig4}     

We formed model spectra using the OSTAR2002 grid
from Lanz \& Hubeny (2003).  These are based on the line-blanketed
and non-LTE atmosphere code TLUSTY and spectrum synthesis code SYNSPEC, 
and the model spectra we used have solar metallicity and an 
adopted microturbulent velocity of 10 km~s$^{-1}$. 
We made a bilinear interpolation in $(T_{\rm eff}, \log g)$
to derive a flux model, and then the fluxes were rebinned on 
the $\log \lambda$ grid of the STIS observations.  
The model spectra were convolved with a rotational broadening 
function using an assumed value of projected rotational velocity 
$V \sin i$ and a linear limb darkening coefficient from the 
tables of Wade \& Rucinski (1985).  

We began by exploring the photospheric line broadening in the 
spectra of Aa1 and Ab.  We derived a $V \sin i = 0$ model spectrum, 
and then compared rotationally convolved versions of the model 
over a grid of $V \sin i$ values and for series of $\approx 50$ \AA\
regions across the spectrum that were free of wind features. 
The resulting $V \sin i$ value and its standard deviation are 
reported in Table~3 for components Aa1 and Ab.  We also list there 
an estimate of $V \sin i$ for Aa2 based on the width of the its CCF peak 
in Figure~2.  Our $V \sin i$ values for Aa1 and Ab are $\approx 13\%$ 
larger than those given by Shenar et al.\ (2015), because Shenar et al.\
attributed part of the line broadening to macroturbulence.  
Component Ab is clearly a rapidly rotating star, and its spectral 
line profiles are broad and shallow.  Furthermore, its flux ratio
in the $V$-band is $F_\lambda ({\rm Ab}) / F_\lambda ({\rm Aa}) = 0.28$ 
(Mason et al.\ 2009), so its flux contribution cannot be ignored 
in radial velocity studies of ground-based spectroscopy.  
These factors are the main reasons why it is has been so difficult
to sort out the different component contributions in the past 
(Mayer et al.\ 2010). 

\placetable{tab3}      

We next considered the best fit $T_{\rm eff}$ models that match 
the UV spectra of Aa1 and Ab.  We formed a grid of models around
the expected temperatures and convolved these with the respective
rotational broadening functions.   Because the main luminosity 
criteria in the UV spectra involve the wind lines and the TLUSTY
models do not account for the wind, we chose to set the gravity 
$\log g$ 
(the atmospheric parameter related to radius) 
to the estimates given by Shenar et al.\ (2015) for $\log g_{2/3}$ 
(the gravity at optical depth $\tau_{\rm Ross}=2/3$). 
The model spectra were rectified in the same way as the observations
by dividing by a second order polynomial fit of the 
relatively line-free regions.  
We formed a $\chi^2$ statistic of the observed and model spectral 
differences over those regions of the spectra free of wind lines, 
and then we estimated the temperature that minimized $\chi^2$. 
Our temperature results appear in Table~3.
Although the derived values of $T_{\rm eff}$ are similar for 
both Aa1 and Ab, we find that Ab is the hotter star by a small margin. 
Our estimates are about $7\%$ hotter than those found by Shenar et al.\ 
(2015), but there are several explanations for this difference.   
First, Shenar et al.\ used optical lines of \ion{He}{1} and \ion{He}{2}
as the primary temperature diagnostics, while our results are based 
on the ultraviolet metallic lines.  Heap et al.\ (2006) found that 
$T_{\rm eff}$ estimates from optical lines led to lower temperatures 
than those from UV lines in their analysis of some O-star spectra, so 
Aa1 and Ab may represent additional cases that show a systematic offset. 
Second, the $T_{\rm eff}$ result is sensitive to the adopted gravity. 
For example, a smaller $\log g$ will result in a lower $T_{\rm eff}$ 
in order to match the observed ionization ratios of the lines. 
If we had adopted lower values from Shenar et al.\ of 
$\log g = \log g_{\rm eff} = \log g_\star (1 - \Gamma)$,
i.e., the gravity corrected for the outward radiative force, 
then our resulting $T_{\rm eff}$ would be closer to their values.
Third, the UV metallic line strengths will increase with increasing 
microturbulent velocity (Heap et al.\ 2006).   Shenar et al.\
derived microturbulent velocities of $\xi_{\rm ph} = 20\pm 5$ and 
$10\pm 5$ km~s$^{-1}$ for Aa1 and Ab, respectively, so it is possible 
that we arrived at a hotter $T_{\rm eff}$ for Aa1 by incorrectly adopting 
$\xi_{\rm ph} = 10$ km~s$^{-1}$ in a solution that relied heavily 
on the strengths of \ion{Fe}{5} lines that locally increase in 
strength with higher $T_{\rm eff}$. 
Fourth, the model spectra are based on the assumption of spherical 
stars, whereas Aa1 and especially Ab will have an oblate shape
due to their rotation.  The local temperature, gravity and 
specific intensity will decrease from pole to enlarged equator,
so that the UV results may preferentially represent conditions 
closer to the hotter poles than those from optical spectra.

The final model spectra are shown as solid lines in Figures 3 and 4. 
The fit is quite satisfactory across the UV spectrum with the 
obvious exception of the strong stellar wind lines in Aa1 
that are absent in the TLUSTY models.  Shenar et al.\ (2015) have made
good fits of the UV wind lines of Aa1 using their PoWR atmospheres code. 
The wind features of Ab, on the other hand, appear similar to the 
photospheric predictions, which probably indicates that Ab is a
lower luminosity star (dwarf or giant). 

 
\section{Summary}                                   

The exquisite angular resolving power of STIS has allowed us
to separate the spectra of $\delta$~Ori~Aa and Ab for the first 
time.   The results verify that Ab is indeed a hot, rapidly 
rotating star that is a significant flux contributor across 
the spectrum.  It will be important to continue angular measurements
of its orbit in order to determine its mass and evolutionary state. 
Component Aa2 in the close binary is relatively faint, but 
we think we have detected its spectral signature in the CCF 
of the difference spectrum that removes the flux of Aa1 (Fig.~2). 
If so, then our mass results are in broad agreement with the 
low mass models of the binary presented by Pablo et al.\ (2015)
and with the Aa1 mass estimated from the observed apsidal motion 
and probable age (Benvenuto et al.\ 2002; Pablo et al.\ 2015). 
If the stars Aa1 and Aa2 are in synchronous rotation with the orbit 
(perhaps at periastron), then the ratio of radii should be 
similar to the ratio of $V \sin i$ values (Table~3), $R_2/R_1 \approx 0.28$. 
Thus, we expect that the flux ratio in the visible part of the 
spectrum will be  
$${{F_\lambda({\rm Aa2})}\over {F_\lambda({\rm Aa1})}} =  
{{f_\lambda({\rm Aa2})}\over {f_\lambda({\rm Aa1})}} 
\left({R_2 \over R_1}\right)^2$$
or $F_\lambda({\rm Aa2}) / F_\lambda({\rm Aa1}) = 0.04$
in the $V$-band for a surface flux ratio of 
$f_\lambda({\rm Aa2}) / f_\lambda({\rm Aa1}) = 0.52$ from
the TLUSTY models for the parameters given in Table~3. 
Detecting the spectrum of Aa2 is challenging, but 
because it may present lines that are much narrower than 
those of Aa1 and Ab, it may be possible to separate out 
all the components in high S/N spectroscopy.  With the stellar 
parameters of Aa1 and Ab now better established, we can reliably 
predict how the blending influence of these components
will appear in the composite line profiles.  By subtracting 
these components, the spectrum of Aa2 may be finally revealed, 
its radial velocity curve measured, and hence accurate masses determined. 
 
 
\acknowledgments 
 
We are grateful to Charles Proffitt and Denise Taylor of STScI 
for their aid in planning the observations with {\it HST}.
We thank Brian Mason, William Hartkopf, and Andrei Tokovinin for 
information about their speckle observations of $\delta$~Ori, and we 
also thank Tomer Shenar for sharing results in advance of publication. 
Support for program \#13450 was provided by NASA through a grant 
from the Space Telescope Science Institute, which is operated by 
the Association of Universities for Research in Astronomy, Inc., 
under NASA contract NAS 5-26555.  
NDR gratefully acknowledges his CRAQ (Qu\'{e}bec) fellowship.
AFJM is grateful for financial aid to NSERC (Canada) and FRQNT (Qu\'{e}bec).

{\it Facilities:} \facility{HST} 
 
 

\clearpage



\clearpage


 
\begin{deluxetable}{ccccc}
\tabletypesize{\scriptsize} 
\tablewidth{0pc} 
\tablenum{1} 
\tablecaption{Radial Velocity Measurements\label{tab1}} 
\tablehead{ 
\colhead{System}           & 
\colhead{Date}             & 
\colhead{Orbital}          & 
\colhead{$V_r$}            & 
\colhead{$\sigma$}         \\  
\colhead{Component}        & 
\colhead{(HJD--2,400,000)} & 
\colhead{Phase}            & 
\colhead{(km s$^{-1}$)}    & 
\colhead{(km s$^{-1}$)}    
} 
\startdata 
Aa1 & 57022.1644 & 0.725   & \phn\phs  99.5  & \phn   1.6 \\
Aa1 & 57064.0356 & 0.029   & \phn    $-85.0$ & \phn   1.5 \\
Aa2 & 57022.1644 & 0.725   &        $-213.3$ &       20.0 \\
Aa2 & 57064.0356 & 0.029   & \phs     256.7  &       20.0 \\
Ab  & 57022.1757 & \nodata & \phn\phs  19.1  & \phn   2.7 \\
Ab  & 57064.0764 & \nodata & \phn\phs  15.9  & \phn   2.9 \\
\enddata 
\end{deluxetable}
  
 
\begin{deluxetable}{lc} 
\tablewidth{0pc} 
\tablenum{2} 
\tablecaption{Orbital Elements for $\delta$~Ori~Aa\label{tab2}} 
\tablehead{ 
\colhead{Element} & 
\colhead{Value}   } 
\startdata 
$P$~(days)                      \dotfill & 5.732436\tablenotemark{a}  \\ 
$T$ (HJD--2,400,000)            \dotfill & 57040.938\tablenotemark{b} \\ 
$e$                             \dotfill & 0.1124\tablenotemark{b}    \\
$\omega$ (deg)                  \dotfill & 144.2\tablenotemark{b}     \\
$K_1$ (km s$^{-1}$)             \dotfill & $104.6 \pm 1.6$            \\ 
$K_2$ (km s$^{-1}$)             \dotfill & $266 \pm 20$               \\ 
$\gamma_1$ (km s$^{-1}$)        \dotfill & $21.1 \pm 1.6$             \\ 
$\gamma_2$ (km s$^{-1}$)        \dotfill & $-14 \pm 20$               \\ 
$M_1$ ($M_\odot$)               \dotfill & $23.3 \pm 3.1$\tablenotemark{c} \\ 
$M_2$ ($M_\odot$)               \dotfill & $9.1  \pm 1.0$\tablenotemark{c} \\ 
$a$ ($R_\odot$)                 \dotfill & $43.0 \pm 2.4$\tablenotemark{c} \\ 
\enddata 
\tablenotetext{a}{Fixed with the value from Mayer et al.\ (2010).}
\tablenotetext{b}{Fixed with values from Pablo et al.\ (2015).}
\tablenotetext{c}{For $i=76\fdg4$ from Pablo et al.\ (2015).}
\end{deluxetable} 

 
\begin{deluxetable}{lccc} 
\tablewidth{0pc} 
\tablenum{3} 
\tablecaption{Stellar Parameters for $\delta$~Ori~A\label{tab3}} 
\tablehead{ 
\colhead{Parameter} & 
\colhead{Aa1}       & 
\colhead{Aa2}       & 
\colhead{Ab} } 
\startdata 
$T_{\rm eff}$ (kK)      \dotfill  & $30.9 \pm 1.8$  & $\approx 23$ & $31.3 \pm 1.7$ \\ 
$\log g$ (cgs)          \dotfill  & 3.37\tablenotemark{a} & 4.0    & 3.5\tablenotemark{a}  \\ 
$V\sin i$ (km s$^{-1}$) \dotfill  & $144 \pm 8$     & $\approx 40$ & $252 \pm 13$ \\ 
$M$ ($M_\odot$)         \dotfill  & $23.3 \pm 3.1$  & $9.1\pm1.0$  & \nodata   \\
\enddata 
\tablenotetext{a}{Fixed with values from Shenar et al.\ (2015).}
\end{deluxetable} 

\end{document}